\documentclass[aps,prl,twocolumn,superscriptaddress]{revtex4-1}
\usepackage{graphicx}
\begin{document}


\title{Coexistence of multiple charge-density waves and superconductivity in SrPt$_2$As$_2$ revealed by $^{75}$As-NMR/NQR and $^{195}$Pt-NMR}


\author{Shinji Kawasaki}
\affiliation{Department of Physics, Okayama University, Okayama 700-8530, Japan}
\affiliation{Research Center of New Functional Materials for Energy Production, Storage and Transport, Okayama University, Okayama 700-8530, Japan}

\author{Yoshihiko Tani}
\affiliation{Department of Physics, Okayama University, Okayama 700-8530, Japan}

\author{Tomosuke Mabuchi}
\affiliation{Department of Physics, Okayama University, Okayama 700-8530, Japan}

\author{Kazutaka Kudo}
\affiliation{Department of Physics, Okayama University, Okayama 700-8530, Japan}
\affiliation{Research Center of New Functional Materials for Energy Production, Storage and Transport, Okayama University, Okayama 700-8530, Japan}

\author{Yoshihiro Nishikubo}
\affiliation{Department of Physics, Okayama University, Okayama 700-8530, Japan}

\author{Daisuke Mitsuoka}
\affiliation{Department of Physics, Okayama University, Okayama 700-8530, Japan}

\author{Minoru Nohara}
\affiliation{Department of Physics, Okayama University, Okayama 700-8530, Japan}
\affiliation{Research Center of New Functional Materials for Energy Production, Storage and Transport, Okayama University, Okayama 700-8530, Japan}

\author{Guo-qing Zheng}
\affiliation{Department of Physics, Okayama University, Okayama 700-8530, Japan}
\affiliation{Research Center of New Functional Materials for Energy Production, Storage and Transport, Okayama University, Okayama 700-8530, Japan}
\affiliation{Institute of Physics and Beijing National Laboratory for Condensed Matter Physics, Chinese Academy of Sciences, Beijing 100190, China}

\begin{abstract}
The relationship between charge density wave (CDW) orders and superconductivity in arsenide superconductor SrPt$_2$As$_2$ with $T_c$ = 5.2 K which crystallizes in the CaBe$_2$Ge$_2$-type structure was studied by $^{75}$As nuclear magnetic resonance (NMR) measurements up to 520 K, and $^{75}$As nuclear quadrupole resonance (NQR) and $^{195}$Pt-NMR measurements down to 1.5 K. At high temperature, $^{75}$As-NMR spectrum and nuclear spin relaxation rate ($1/T_1$)  have revealed two distinct CDW orders, one realized in the As-Pt-As layer below $T_{\rm CDW}^{\rm As(1)}$ $=$ 410 K and the other in the Pt-As-Pt layer below $T_{\rm CDW}^{\rm As(2)}$ $=$ 255 K. The $1/T_1$ measured by $^{75}$As-NQR shows a clear Hebel-Slichter peak just below $T_c$ and decreases exponentially well below $T_c$. Concomitantly, $^{195}$Pt Knight shift decreases below $T_c$. Our results indicate that superconductivity in SrPt$_2$As$_2$ is  in the spin-singlet state with an $s$-wave gap and is robust under the two distinct CDW orders in different layers.
\end{abstract}

\pacs{}

\maketitle


Since the discovery of the iron-arsenide superconductor LaFeAsO$_{1-x}$F$_x$\cite{Kamihara}, the relationship between a spin-density wave and/or a structure instability and the high transition temperature ($T_c$) superconductivity has become a hot topic\cite{Kamihara,Oka,Zhou,Ueshima,Baek}. Previous extensive studies on arsenides \cite{Alloul,LFang,Imai,Nakai,Tabuchi,Kaminski,Pratt} have shown that  $T_c$ and the antiferromagnetic spin fluctuation due to a Fermi-surface nesting among multiple electronic bands correlate with each other as predicted theoretically\cite{Kuroki,Mazin,Chubkov}.  The charge density wave (CDW) order is another ground state derived from a Fermi-surface nesting, and can compete and/or coexist with superconductivity\cite{Gabovich}.  Therefore, investigation of the relationship between CDW and  superconductivity may also shed light on the understanding of the properties of the Fe pnictides.

The Pt-arsenide superconductor SrPt$_2$As$_2$ ($T_c$ = 5.2 K\cite{Kudo}) crystallizes in the CaBe$_2$Ge$_2$-type structure. Different from the ThCr$_2$Si$_2$-type structure,  SrPt$_2$As$_2$ has two inequivalent As and Pt sites due to the stacking of As-Pt-As and Pt-As-Pt layers along the $c$-axis\cite{Imre}.  Interestingly, the structural phase transition and the formation of one dimensional CDW with a modulation vector $\vec q_{\rm CDW}$ = 0.62 $\vec a^\star$ = (0.62, 0, 0) in the As-Pt-As layer at $T_{\rm CDW}$ $\sim$ 450 K were found by X-ray and transmission electron microscopy measurements\cite{Imre,Fang,Wang}.  In particular, it has been suggested that the multiple bands formed by the Pt 5$d$ state in the As-Pt-As layer plays a major role to induce the CDW order\cite{Imre,Kim}. However, since $T_{\rm CDW}$ is much higher than room temperature, the experimental result on $T_{\rm CDW}$ has been limited to the resistivity measurement\cite{Fang}.  Thus, the relationship between the CDW and  the superconducting order in SrPt$_2$As$_2$ is still unknown.

In this Rapid Communication, we report $^{75}$As-NMR/NQR and $^{195}$Pt-NMR studies in a wide temperature range from 1.5 to 520 K using a polycrystalline SrPt$_2$As$_2$ sample to address the above issue.  Above $T_c$, the  $^{75}$As-NMR spectrum and the nuclear-spin lattice relaxation rate ($1/T_1$) indicate that there are two CDW transitions; one is the CDW order below $T_{\rm CDW}^{\rm As(1)}$ = 410 K in the As-Pt-As layer and the other below $T_{\rm CDW}^{\rm As(2)}$ = 255 K in the Pt-As-Pt layer. Below $T_{\rm CDW}^{\rm As(2)}$, we find that 1/$T_1T$ becomes constant down to $T_c$. In the superconducting state, 1/$T_1$ measured by the $^{75}$As NQR at the CDW ordered site shows a clear Hebel-Slichter peak and decreases exponentially well below $T_c$, evidencing the coexistence of the CDW order and the superconductivity.  Concomitantly, $^{195}$Pt Knight shift decreases below $T_c$, indicating that the superconductivity in SrPt$_2$As$_2$ is in the spin-singlet state with an $s$-wave gap.

The polycrystalline SrPt$_2$As$_2$ sample for the NMR/NQR measurements was prepared by a solid-state reaction method as reported elsewhere\cite{Kudo}. NMR/NQR measurements were carried out by using a phase-coherent spectrometer. For $^{75}$As-NMR measurements above 300 K, a homemade air-cooled variable temperature insert was used. The $^{195}$Pt-NMR spectrum was measured around $T_c(H)$ at the fixed field of $H$ = 0.505 T. The values of $T_c$ at $H$ = 0 and 0.505 T was determined by the ac-susceptibility measurement using NMR coil. NMR/NQR spectra were taken by changing rf frequency and recording spin echo intensity step by step. Due to strong anisotropy\cite{Fang}, the powdered sample was easily oriented to the direction of $H \parallel ab$ plane\cite{Grafe} at a high field of 12.951T. We take this advantage to perform $^{75}$As-NMR measurements. The $T_1$ above $T_c$ was measured at the center peak ($m = 1/2\leftrightarrow-1/2$). Below $T_c$, $^{75}$As NQR ($H$ = 0) was utilized for $1/T_1$ measurements. The $T_1$ was measured by using a single saturating pulse, and is determined by fitting the recovery curve of the nuclear magnetization to the theoretical functions: $1-M(t)/M_0$ = $0.9\exp$$(-6t/T_1)$+$0.1\exp(-t/T_1)$ in NMR and  $1-M(t)/M_0$ = $\exp(-3t/T_1)$ in NQR \cite{Andrew,Narath}, where $M_0$ and $M(t)$ are the nuclear magnetization in the thermal equilibrium and at a time $t$ after the saturating pulse.

The CDW orders were probed through the nuclear quadrupole interaction $\mathcal{H}_{\rm Q}$ = $(h \nu_{\rm Q}/6)(3{I_z}^2-I(I+1))$\cite{Blinc,Berthier,Abragam}. Here, $\nu_{\rm Q}$ is the NQR frequency which measures the electric field gradient (EFG) generated by a carrier distribution and a lattice contribution originate from CDW order. For NMR, the total Hamiltonian is a sum of $\mathcal{H}_{\rm Q}$ and the Zeeman term as $\mathcal{H}$ = $\gamma$$\hbar$$\overrightarrow{I}\cdot\overrightarrow{\textit{H}_{0}}$  + $\mathcal{H}_{\rm Q}$. In the case of the external field $H_0$ being perpendicular to the principal axis of $\nu_{\rm Q}$, the NMR center peak frequency shifts from $f_{0}$ = $\gamma_{\rm N}$$H_0$ to $f_{1}$ = $K_{\perp}$$\gamma_{\rm N}$$H_0$ + $3\nu_{\rm Q}^2/16(1+K_{\perp})\gamma_{\rm N}H_0$, where $K_{\perp}$ is the Knight shift in the direction perpendicular to the principal axis of $\nu_{\rm Q}$\cite{Abragam}. For arsenides, $K_{\perp}$ at As sites corresponds to Knight shift in the $ab$ plane\cite{Grafe,Matano}.  

As a result of one-dimensional CDW order, the wave modulation causes a spatial distribution in the NQR frequency as $\nu$ = $\nu_{\rm Q}$+$\nu_1$$\cos$$[\phi(x)]$ where $\nu_1$ corresponds to the amplitude of the modulation wave. The density of the spectral line shape strongly depends on the spatial variation of $\phi(x)$ as $f(\nu)$ $=$ $1/(2\pi d \nu / d \phi)$\cite{Blinc}. For incommensurate order, $\phi(x)$ increases linearly with $x$ (plane wave). In this case, the spectral line shape shows an edge singularity at $\nu$ = $\nu_{\rm Q}$ $\pm$ $\nu_1$ as $f(\nu)$ = 1/(2$\pi$$\nu_1$$(1-X^2$)$^{1/2}$) where $X$ = ($\nu-\nu_{\rm Q}$)/$\nu_1$. On the other hand, if CDW and a crystal lattice correlate with each other, soliton excitations\cite{Rice} and/or discommensuration\cite{McMillan} will occur and $\phi(x)$ becomes nonlinear in $x$ (solitonic wave).  In this case, a commensurate region appears and the spectral line shape shows a peak at $\nu=\nu_{\rm Q}$ in addition to the edge singularity at $\nu_{\rm Q}$ $\pm$ $\nu_1$\cite{Blinc}. The NMR/NQR spectrum $F(\nu)$ in the CDW ordered state is given by $F(\nu)=\int{P}(\nu)f(\nu)d\nu$ with the parameters of $\nu_1$ and $\nu_{\rm Q}$, where $P(\nu)$ is a distribution function. Thus, one obtains two (three) peak structures for the plane (solitonic) wave order in the NMR/NQR spectrum. Here, we employed Lorentzian function as $P(\nu)$.  Furthermore, the temperature dependence of 2$\nu_1$ corresponds to the order parameter of the CDW order\cite{Blinc}.

 \begin{figure}
 \begin{center}
 \includegraphics[width=7.5cm]{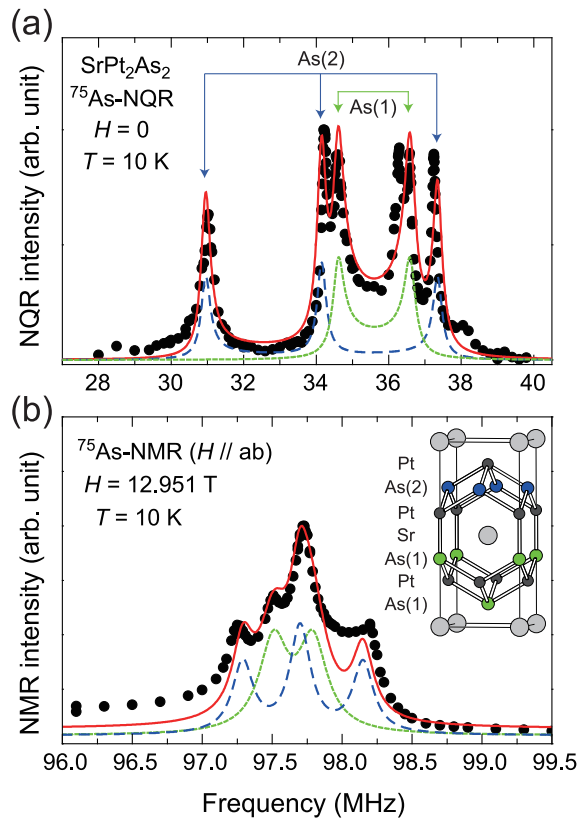}
 \end{center}
 \caption{(Color online)  $^{75}$As NQR and NMR  spectra. The dotted and dashed lines are the simulated spectra assuming two different CDWs using same parameters (see text), respectively. The solid lines are the sum of the simulated spectra. }
 \label{f1}
 \end{figure}

Figures 1(a) and 1(b) show the spectra of the $^{75}$As-NQR and the $^{75}$As-NMR center peak obtained at $T$ = 10 K, respectively, along with an illustration of the crystal structure of SrPt$_2$As$_2$ in the inset of Fig. 1(b). Here, we denote two As sites as As(1) and As(2) in the As-Pt-As and the Pt-As-Pt layers, respectively. The obtained spectra consist of multiple peaks in both NQR and NMR although there are only two As sites in SrPt$_2$As$_2$ with the undistorted CaBe$_2$Ge$_2$-type structure. This is clearly due to the addition of $\nu_1$ at the As-sites below $T_{\rm CDW}$. As shown in Figs. 1(a) and 1(b), we find that a combination (solid lines) of a plane-wave order [dotted line denoted as As(1)] and a solitonic-wave order [dashed line denoted as As(2)] can reproduce  reasonably well both spectra at the same time. This result indicates that there are two different CDW orders in the ground state of SrPt$_2$As$_2$.  The obtained NQR and NMR parameters for both sites are summarized in Table I. 

\begin{table}[h]
 \caption{$^{75}$As-NQR/NMR parameters for SrPt$_2$As$_2$ obtained at 10 K.}
  \begin{center}
  \begin{tabular}{|c|c|cc|c|}
    
    \hline
    
        As site        & $T_{\rm CDW}$(K)    & $\nu_{\rm Q}$(MHz)   & $\nu_1$(MHz)  & $K_{\rm ab}$(\%) \\
   \hline
    \hline
      As(1) &  410            & 35.60   &  1.00  & 0.75 \\
    \hline
     As(2)    & 255         &  34.15  &  3.20  & 1.02 \\

    \hline  
  \end{tabular}
 \end{center}
\end{table}

\begin{figure}
\begin{center}
\includegraphics[width=8.5cm]{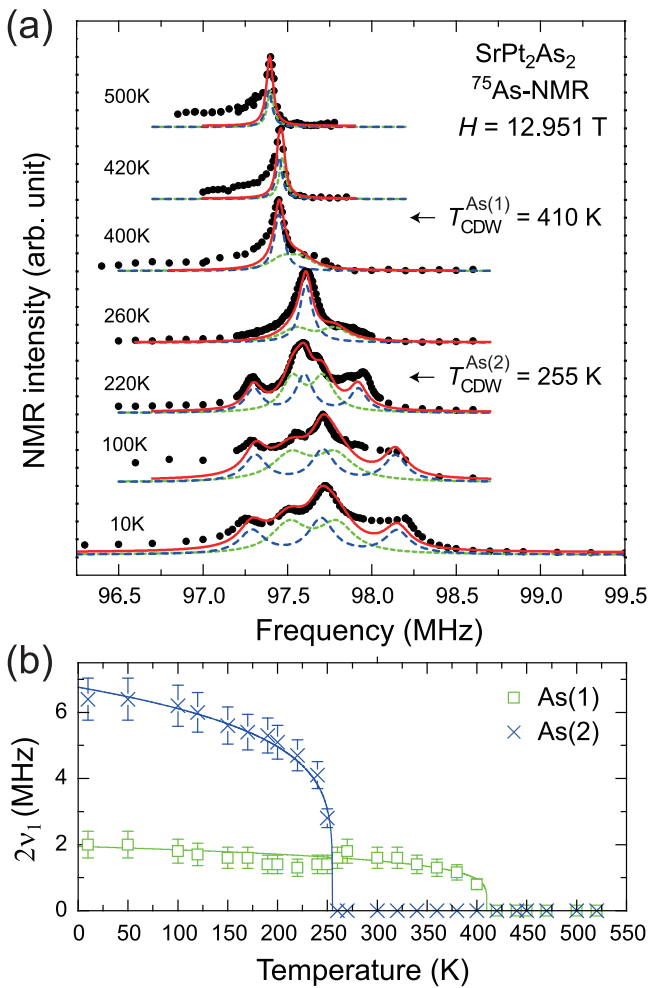}
\end{center}
\caption{(Color online) (a) Temperature dependence of the $^{75}$As-NMR center peak for $H \parallel ab$ plane obtained at $H$ = 12.951 T. The dotted and dashed lines are the calculated curves for As(1) and As(2), respectively (see text). The solid lines are the sum of the dotted and dashed curves. (b) Temperature dependence of $2\nu_1$ for both As(1) and As(2). The solid curves are the fittings of 2$\nu_1$ $\propto$($T_{\rm CDW}-T$)$^\beta$. Error bars are deduced from a combination of systematic error and standard deviation.} 
\label{f2}
\end{figure}

 Next, we measured the $^{75}$As-NMR spectrum up to 520 K and obtained the temperature dependence of $2\nu_1$  for both sites by fitting.  Figure 2(a) shows the temperature dependence of the NMR spectrum along with the simulation. Above $T$ = 420 K, a sharp peak is observed which is consistent with the absence of $\nu_{1}$ above $T_{\rm CDW}$. Here, the tail structure of the lower frequency side of the spectrum comes from unoriented powders. Below $T$ = 400 K, a part of the spectrum shifts to the higher frequency and becomes broad, indicating that the CDW order occurs. Notably, the main peak still exists even though the CDW order sets in. As shown by the fittings, this indicates that $\nu_1$ exists at one As site [As(1)] but is absent at the other site [As(2)]. This result is consistent with the previous reports which suggest one dimensional CDW order only in the As-Pt-As layer\cite{Imre,Fang}. Thus, we assigned that As(1) is in the As-Pt-As layer and As(2) is in the Pt-As-Pt layer.  Furthermore, below $T$ = 260 K, we find that the $^{75}$As-NMR spectrum changes its shape drastically. From the simulation, we suggest that $\nu_{1}$ also appears at As(2) with larger amplitude than that at As(1) and, the CDW order also occurs in the Pt-As-Pt layer below $T$ = 260 K. As shown by the solid lines in Fig. 2(a), the temperature dependence of the $^{75}$As-NMR spectrum can be reproduced by assuming two CDW orders as determined in Fig. 1.  Note that the value of 2$\nu_1$ is determined purely by the distance between the edge singularity of the spectrum, being independent of the selected values of $\nu_{\rm Q}$ and $K_{\rm ab}$. 

 Figure 2(b) shows the temperature dependence of 2$\nu_1$ for both As(1) and As(2). As shown by the solid lines in Fig. 2(b), from the formula 2$\nu_1$ $\propto$ ($T_{\rm CDW}-T$)$^\beta$, we obtain $T_{\rm CDW}^{\rm As(1)}$ = 410 K with $\beta$ = 0.20, and $T_{\rm CDW}^{\rm As(2)}$ = 255 K with $\beta$ = 0.19.  From Fig. 2, we conclude that, different from previous reports\cite{Imre,Fang}, the solitonic wave order occurs in the Pt-As-Pt layer at $T_{\rm CDW}^{\rm As(2)}$ = 255 K in addition to the plane wave order in the As-Pt-As layer at  $T_{\rm CDW}^{\rm As(1)}$ = 410 K in SrPt$_2$As$_2$.     
 
 \begin{figure}
 \begin{center}
 \includegraphics[width=8cm]{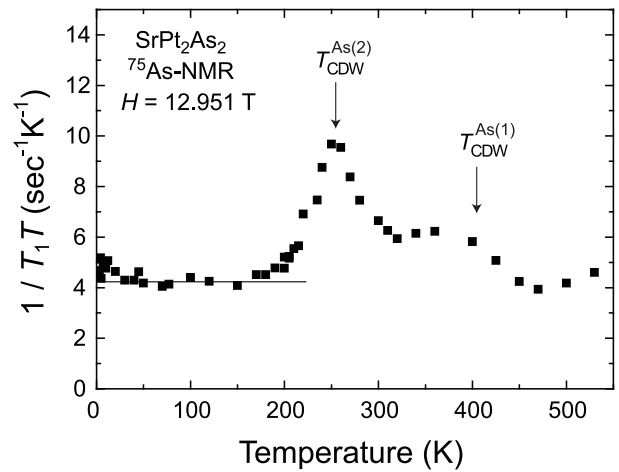}
 \end{center}
 \caption{Temperature dependence of the $1/T_1T$ measured at $H$ = 12.951 T. The solid arrows indicate $T_{\rm CDW}^{\rm As(1)}$ = 410 K and $T_{\rm CDW}^{\rm As(2)}$ = 255 K, respectively. The solid line indicates the relation $1/T_1T$ = const. } 
 \end{figure}

The above conclusion is further supported by the $1/T_1$ results. Figure 3 shows the temperature dependence of the quantity 1/$T_1T$, which shows peaks at $T_{\rm CDW}^{\rm As(1)}$ = 410 K and $T_{\rm CDW}^{\rm As(2)}$ = 255 K.   Notably, the amplitude of $1/T_1T$ at $T_{\rm CDW}^{\rm As(2)}$ is larger than that at $T_{\rm CDW}^{\rm As(1)}$ = 410 K. This is consistent with the result that the amplitude of $\nu_1$ at As(2) is larger than that at As(1). Although it is difficult to clarify the origin of these peaks quantitatively in the present stage, they probably originated from the fluctuations of EFG\cite{Obata} due to the CDW orders.  Below $T$ = 150 K, $1/T_1T$ becomes a constant, indicating a Fermi liquid background for superconductivity in SrPt$_2$As$_2$.  The observed result is completely different from iron arsenides where $1/T_1T$ increases significantly with decreasing $T$ due to antiferromagnetic spin fluctuation associated with Fermi-surface nesting\cite{Oka,Imai,Nakai}.

 \begin{figure}
 \begin{center}
 \includegraphics[width=7.5cm]{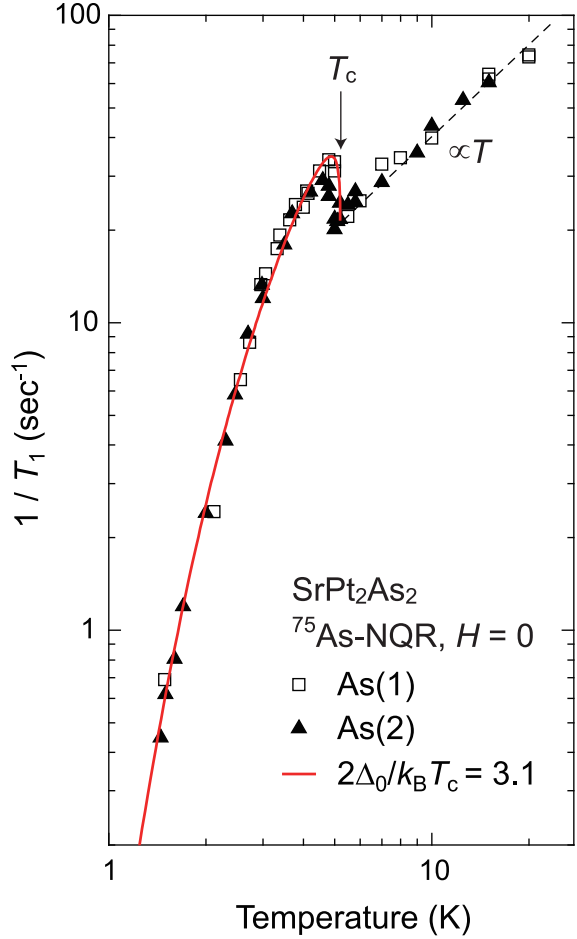}
 \end{center}
 \caption{(Color online) Temperature dependence of $1/T_1$ for both As(1) (open square) and As(2) (solid triangle) sites. The solid curve is the calculated result assuming a single $s$-wave gap. The dotted line indicates the relation $1/T_1T$ = const.  }
 \label{f4}
 \end{figure}

Finally, we turn to the superconducting state. Figure 4 shows the temperature dependence of the $1/T_1$ for both As(1) and As(2) sites. The site dependence of the $T_1$ value is negligible.   As seen in the figure, 1/$T_1$ shows a clear Hebel-Slichter peak just below $T_c$ = 5.2 K and decreases exponentially, which are characteristics of conventional BCS superconductors\cite{Masuda}.  This result also evidences homogeneous coexistence of two CDW orders and superconductivity because $T_1$ is measured at the CDW ordered sites of As. The solid curve in Fig. 4 is a calculation using the BCS model. Here, the relaxation rate below $T_{\rm c}$ ($1/T_{1s}$) is expressed as

\begin{eqnarray*}
\frac{T_1(T_c)}{T_{1s}}= \frac{2}{k_BT_c} \int [N_s(E)^2 + M_s(E)^2] f(E)\left[1-f(E) \right] dE,
\end{eqnarray*}

where $M_s(E) = \frac{N_0\Delta}{\sqrt{E^2- \Delta^2}}$ is the anomalous density of states due to the coherence factor,  $N_s(E) = \frac{N_0E}{\sqrt{E^2- \Delta^2}}$ is the density of states in the superconducting state, $\Delta$ is the energy gap, $N_0$ is the density of states in the normal state, and $f(E)$ is the Fermi distribution function. We convolute $M_s(E)$ and $N_s(E)$ with a broadening function assuming a rectangle shape with a width 2$\delta$ and a height 1/2$\delta$\cite{Hebel}. The solid curve in Fig. 4 is the calculated result with the parameters of 2$\Delta(0)/k_{\rm B}T_{\rm c}$ = 3.1  and $r = \Delta (0) / \delta = 4$. The value of 2$\Delta(0)/k_{\rm B}T_{\rm c}$ is comparable to that of the BCS weak coupling limit. Our results strongly suggest that the superconducting gap of SrPt$_2$As$_2$ fully and equally opens among the multiple Fermi surfaces. This is clearly different from multigap superconductivity observed in iron  arseneides\cite{Oka}. The present result is also different from the previous specific heat measurement\cite{Xu} which suggested that two different gaps open on different Fermi surfaces.

To elucidate the spin symmetry of the Cooper pairs, we measured the $^{195}$Pt Knight shift at $H$ = 0.505 T [$T_c(H)$ $\sim$ 4.2 K].  Figure 5 inset shows the $^{195}$Pt-NMR spectra. It is clearly seen that the line width becomes broad below $T_c$, due to the distribution of the magnetic field in the superconducting state\cite{Redfield}. Most importantly, the peak position shifts to a lower frequency i.e. Knight shift decreases below $T_c$. Figure 5 shows the temperature dependence of the $^{195}$Pt Knight shift deduced from the peak position with $^{195}$$\gamma_{\rm N}$ = 9.094 MHz T$^{-1}$. The obtained Knight shift clearly decreases below $T_c$, evidencing the spin-singlet pairing.

\begin{figure}
 \begin{center}
 \includegraphics[width=8cm]{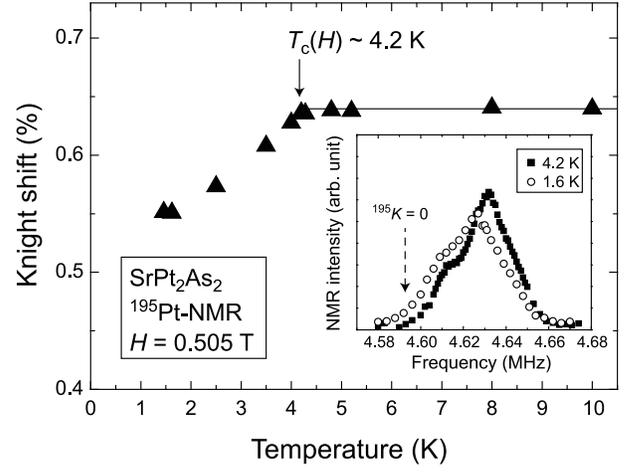}
 \end{center}
 \caption{Temperature dependence of the $^{195}$Pt Knight shift measured at $H$ = 0.505 T. The solid line is an eye guide. The inset shows the spectrum above and below $T_c$, respectively. The dotted arrow indicates the  position at $^{195}K$ = 0.   }
 \label{f5}
 \end{figure}

In conclusion, we have presented the extensive $^{75}$As-NMR/NQR and $^{195}$Pt-NMR results on Pt-arsenide SrPt$_2$As$_2$. We have shown that two CDW orders occur in SrPt$_2$As$_2$ at $T_{\rm CDW}^{\rm As(1)}$ = 410 K and $T_{\rm CDW}^{\rm As(2)}$ = 255 K, respectively. From the temperature dependence of the NMR spectrum up to 520 K, it was found that the two CDW orders take place with two different transition temperatures  in two Pt-As layers. The superconductivity occurs in the background of the Fermi-liquid state and coexists with the two CDW orders, in the spin-singlet state with an $s$-wave gap. Our results help us understand the relationship between superconductivity and density wave orders originating from multiple Fermi surfaces in various materials including the iron-arsenide superconductors.

We thank Satoki Maeda for useful discussions on data analysis. This work was supported in part by research grants from MEXT (Grants No. 22103004 and No. 25400374).

\end{document}